# Análise Termodinâmica da aceleração de uma massa

Rodrigo de Abreu
Centro de Electrodinâmica e Departamento de Física
do IST

**Abstract**

We analyse the acceleration of a mass with a simple structure taking into account Thermodynamics. Two situations are analysed. The first one for the application of a localized force to a point of the mass. The second one for the application of a force to the entire mass. The two situations are not equivalent. For the first situation we have an increase of temperature of the mass, resulting from an internal damping, during a transient.

**Resumo**

A necessidade de considerar a termodinâmica em problemas geralmente tratados como puramente mecânicos tem dado origem a diversos artigos. A análise termodinâmica da aceleração de uma massa é sobre esta mesma matéria. Através de duas análises, mostra-se em que condições a aceleração de uma massa extensa, uma caixa com um gás no seu interior, que parte do repouso, necessita de um tratamento termodinâmico. Numa 1ª análise, a aceleração é provocada por uma força aplicada num ponto. A acção desta força altera a distribuição da densidade do gás e origina um aumento de temperatura. Prova-se que este aumento de temperatura dá origem a um aumento de massa do gás. Na 2ª análise considera-se que a aceleração é devida a um campo gravitacional constante. Mostra-se que neste caso a temperatura não varia e desta forma as duas situações analisadas não são equivalentes. Da 1ª análise conclui-se que uma massa submetida a uma força variável no tempo dissipa constantemente energia no seu interior, com aumento de entropia, como se existisse uma força de atrito interno.

### Introdução

A síntese entre mecânica e termodinâmica tem vindo a ser feita em diversos artigos [1-13]. As duas análises que são aqui apresentadas pretendem contribuir para esclarecer, aplicadas a um modelo concreto, aspectos conceptuais delicados: o problema da validade da mecânica do ponto aplicada a um corpo extenso [1]; o problema do significado de massa [14] e o problema da variação da massa própria, apenas como resultado da acção de uma força - sem emissão ou absorção de partículas [15].

Considera-se, em ambas as análises, uma massa acelerada, no vácuo, por uma força $F$ constante. Esta massa é constituída pela massa $M$ de um invólucro rígido, uma caixa, que contém um gás, com $N$ partículas, cada uma das quais com massa $m$.

*Demonstra-se que embora a força que actua sobre a caixa e o gás tenha igual módulo, F, o resultado da acção desta força difere. No 1º caso, a força está aplicada num ponto da caixa, e no 2º a força actua simultaneamente sobre todas as massas que constituem a caixa e o gás.*

Na 1ª análise, a caixa é acelerada a partir do repouso por uma força que actua longitudinalmente, aplicada a uma das suas bases (Fig.1). Esta força que é aplicada instantaneamente, acelera a caixa e altera a distribuição da densidade de partículas no interior que, inicialmente, era uniforme. Determina-se em 1. a distribuição final de equilíbrio que se estabelece no interior da caixa e, em 1.1, calcula-se o aumento de temperatura do gás resultante da aceleração. Este aumento de temperatura está associado a um aumento de energia interna e também, mostra-se em 1.2, a um aumento de massa. Deste modo se a massa for submetida a uma força que varia constantemente e bruscamente de sentido, como quando se agita um tubo de ensaio que contém no seu interior um liquido, dissipa constantemente energia no seu interior, com aumento de entropia e com aumento de massa - existe uma força de atrito interno resultante das colisões das partículas do gás nas bases da caixa. Esta análise mostra que só durante a deformação do gás é que o conjunto caixa-gás não é equivalente a um ponto material [1].

Em 2., numa 2ª análise, considera-se que a mesma caixa está agora suspensa por um fio, na presença de um campo gravitacional constante. O módulo da força exercida pelo fio no estado de equilíbrio inicial é $F$, e portanto o peso da caixa e do gás tem também módulo $F$. A distribuição do gás no interior da caixa no estado de equilíbrio inicial é a mesma que no estado final da 1ª análise. Cortado o fio, a caixa e o gás sofrem a aceleração do campo gravitacional. Mostra-se que neste caso a temperatura do gás não varia e desta forma as duas situações analisadas não são equivalentes. Neste segundo caso, tudo se passa como se de um ponto material se tratasse. Embora exista deformação, o centro de massa do conjunto comporta-se como um ponto material dado o trabalho da força resultante ser a soma dos trabalhos das forças que actuam nos centros de massa dos sub-sistemas gás e caixa [1].

Estas duas análises contradizem algumas ideias correntes, e.g. as expostas em [1].

## 1. A 1ª Análise

Consideremos uma massa submetida a uma força $F = const.$ de acordo com a Fig. 1. A massa da caixa é $M$ e o gás é constituído por $N$ partículas em que cada uma tem massa $m$. A caixa inicialmente está em repouso e o gás está distribuído uniformemente no seu interior. O exterior da caixa é vácuo.

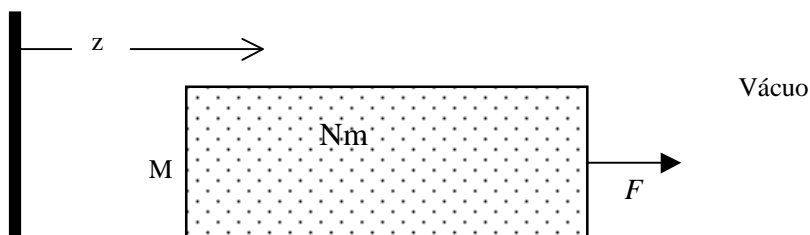

Fig. 1 - A massa da caixa é $M$ e a massa de cada partícula que se encontra no interior da caixa é $m$. O número de partículas é $N$. A força $F$ é aplicada numa das bases da caixa.

Quando se aplica a força *F* o módulo da aceleração $a_{CM}$ do centro de massa do conjunto caixa - gás satisfaz a equação, em qualquer instante,

$$a_{CM} = \frac{F}{(Nm+M)} \qquad (1)$$

De (1)

$$F\Delta z_{CM} = \frac{1}{2}(M+Nm)v_{CM}^2 \qquad (2)$$

em que $z_{CM}$ e $v_{CM}$ são respectivamente a posição e a componente da velocidade do centro de massa. Para um ponto material o centro de massa coincide com o próprio ponto. Se o ponto tiver massa (*M+Nm*), da equação (1), deriva-se, como é bem conhecido, que o trabalho da força que actua sobre o ponto é igual á variação da energia cinética, eq. (2), e, portanto, para o centro de massa de um conjunto de pontos materiais, deriva-se relação idêntica, embora neste caso o produto da força pelo deslocamento do centro de massa não seja o trabalho da força, como se verá adiante.

O produto da força pelo deslocamento do centro de massa não é igual ao trabalho da força *F*, porque o gás no interior da caixa deforma-se e, consequentemente, o centro de massa desloca-se no interior da caixa. Desta forma o deslocamento do centro de massa não coincide com o deslocamento da caixa e portanto o produto da força pelo deslocamento do centro de massa não é igual ao trabalho da força. Este trabalho da força *F* é o produto da força pelo deslocamento da caixa, $\Delta z_{CX}$.

Por outro lado o trabalho da força *F* é igual à variação da energia cinética da caixa e do gás

$$F\Delta z_{CX} = \frac{1}{2}(M+Nm)v_{CM}^2 + \Delta U_{gás}^{int} \qquad (3)$$

dado estarmos a admitir que a variação da energia interna *U* do gás é apenas cinética.

Quando o gás atinge um novo estado de equilíbrio o centro de massa permanece numa posição constante no interior da caixa, passamos a ter

$$F\Delta z_{CX} = \frac{1}{2}(M+Nm)v_{CX}^2 = F\Delta z_{CM} = \frac{1}{2}(M+Nm)v_{CM}^2. \qquad (4)$$

A variação de energia interna do gás apenas se dá durante o regime transitório em que há deformação, ou seja durante o intervalo de tempo em que o centro de massa se desloca no interior da caixa.

Das equações (2) e (3) temos que

$$F(\Delta z_{CX} - \Delta z_{CM}) = Fd' = \Delta U_{gás}^{int} \qquad (5)$$

em que *d´* é a distância percorrida pelo centro de massa durante a deformação no referencial da caixa (Fig.2)

$$d' = (\frac{L}{2} - z_{CM0}). \qquad (6)$$

*L/2* é a posição do centro de massa antes da deformação e $z_{CM0}$ é a posição do centro de massa depois da deformação, no referencial da caixa.

(Note-se que se o corpo estiver em movimento rectilíneo e uniforme e se o gás estiver inicialmente com uma distribuição uniforme e se for aplicada à caixa uma força de módulo *F*, mas com sentido contrário ao movimento, o centro de massa desloca-se de *d´* mas no mesmo sentido do movimento. Deste modo conclui-se que se a força *F* inverter periódicamente de sentido, o centro de massa desloca-se simétricamente em relação ao centro da caixa com aumento de energia interna do gás.)

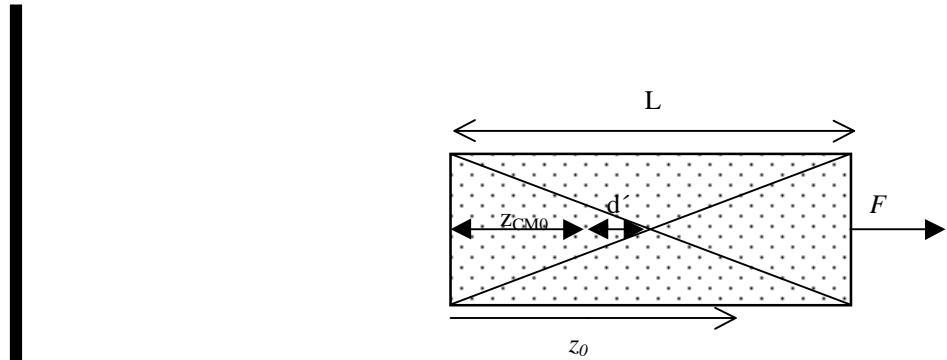

Fig.2 - O centro de massa do gás desloca-se no sentido contrário ao movimento até atingir uma posição de equilíbrio. O centro de massa da caixa que é rígida está fixo em relação ás paredes da caixa e coincide com o centro geométrico, dado estarmos a admitir que a distribuição de massa da caixa é uniforme. $z_{CM0}$ é a ordenada do centro de massa, do conjunto caixa-gás, no referencial da caixa. $z_0$ é a ordenada no referencial da caixa.

A posição do centro de massa no interior da caixa é

$$z_{CM0} = \frac{M\frac{L}{2} + Nmz_{g0}}{M + Nm} \qquad (7)$$

em que $z_{g0}$ é a posição do centro de massa do gás no referencial da caixa. De (6) e (7) temos

$$d' = \frac{Nm}{Nm + M}(\frac{L}{2} - z_{g0}). \qquad (8)$$

Se considerarmos que o gás é monoatómico, de (5), temos

$$F(\Delta z_{CX} - \Delta z_{CM}) = Fd' = \Delta U_{gás}^{int} = N\frac{3}{2}k(T_2 - T_1) \qquad (9)$$

em que $k$ é a constante de Boltzmann e $T_1$ e $T_2$ são respectivamente as temperaturas inicial e final do gás. Da equação (9) temos que a determinação de $d'$ em função da temperatura final $T_2$ permite calcular essa mesma temperatura, o que faremos a seguir.

## 1.1 Determinação da temperatura final do gás

O gás no referencial do centro de massa está submetido a uma aceleração constante de módulo $a_{CM}$ que resulta do centro de massa ter uma aceleração constante de módulo $a_{CM}$. Mas no referencial do centro de massa a aceleração tem sentido contrário ao movimento. É como se o gás estivesse submetido a um campo gravitacional de aceleração $g = a_{CM}$. Desta forma a densidade de partículas do gás no interior da caixa, após o centro de massa deixar de se deslocar em relação á caixa, é dada por

$$n(z_0) = n(0)e^{-\frac{mg}{kT_2}z_0} \qquad (10)$$

em que $z_0$ é a "altura" medida em relação á base da caixa e $n(0)$ é a densidade de partículas para $z_0 = 0$. De (10) determinamos o valor de $n(0)$, dado que o número de partículas no interior da caixa é $N$

$$\int_0^L n(z_0)dz_0 = N = \int_0^L n(0)e^{-\frac{mg}{kT_2}z_0}dz_0 = n(0)\int_0^L e^{-\frac{mg}{kT_2}z_0}dz_0 \qquad (11)$$

$$n(0) = \frac{N}{\int_0^L e^{-\frac{mg}{kT_2}z_0}dz_0} = \frac{N}{(1-e^{-\frac{mg}{kT_2}L})} \times \frac{mg}{kT_2} \qquad (12)$$

O centro de massa do gás no referencial da caixa é dado por

$$z_{g0} = \frac{\int_0^L z_0 \times m \times n(z_0)dz_0}{Nm} \qquad (13)$$

e de (10) e (12), vem

$$z_{g0} = \frac{Le^{-\frac{mgL}{kT_2}}}{e^{-\frac{mgL}{kT_2}}-1} + \frac{kT_2}{mg} \qquad (14)$$

De (8), (9) e (14) temos

$$F \times \frac{Nm}{Nm+M}(\frac{L}{2} - \frac{Le^{-\frac{mgL}{kT_2}}}{e^{-\frac{mgL}{kT_2}}-1} - \frac{kT_2}{mg}) = N\frac{3}{2}k(T_2 - T_1) \qquad (15)$$

De (15), finalmente, determina-se o valor de $T_2$. Para $N$ e $m$ conhecidos, a temperatura $T_2$ é tanto maior quanto maior for $F$ e $L$ e menor $M$. Este efeito de aumento de temperatura também se verifica se o corpo estiver em movimento e for travado por uma força $F$. Neste caso também podemos ter acelerações muito elevadas se o corpo tender para o repouso num curto intervalo de tempo, numa distância muito pequena. Por exemplo se a velocidade da caixa for 300 km/h ou seja 83,3 m/s e variar para zero num centésimo de segundo a aceleração é 833 m/s$^2$. Se variar para zero num milésimo de segundo passa a ser 8330 m/s$^2$ que é uma aceleração da mesma ordem de grandeza da que é necessária para imprimir uma velocidade de 8 km/s num segundo, velocidade necessária para colocar uma massa em órbita da terra, a baixa altitude.

### 1.2 A variação de massa

Nas condições desta 1ª análise vamos demonstrar que a acção da força $F$ implica necessariamente uma variação de massa do gás e portanto uma variação de massa do conjunto gás-caixa. De facto, a teoria da relatividade mostra que a energia e a massa estão associadas através da fórmula de Einstein $E = mc^2$. Para baixas velocidades, esta fórmula é uma boa aproximação do resultado que se obtém através da análise clássica, que na 1ª análise deu o resultado expresso na equação (3). Essa equação pode portanto ser escrita na forma

$$F\Delta z_{CX} = m_2 c^2 - m_1 c^2 = \frac{m_{02}}{(1-\frac{v_{CM}^2}{c^2})^{\frac{1}{2}}}c^2 - m_{01}c^2 \cong \frac{1}{2}(M+Nm)v_{CM}^2 + \Delta U_{gás}^{int} \qquad (16)$$

em que $m_{02}$ e $m_{01}$ representam, respectivamente, a massa do conjunto caixa - gás, em dois estados: após o centro de massa do gás ter atingido a posição de equilíbrio dada por (14) e, portanto, quando as velocidades dos centros de massa da caixa e do gás forem $v_{CM}$; no início da actuação da força $F$ em que a velocidade do centro de massa era zero e portanto $m_1 = m_{01}$. Por outro lado se a velocidade do centro de massa for muito inferior a $c$ podemos escrever

$$F\Delta z_{CX} = \frac{m_{02}}{(1-\frac{v_{CM}^2}{c^2})^{\frac{1}{2}}}c^2 - m_{01}c^2 = \frac{1}{2}m_{02}v_{CM}^2 + m_{02}c^2 - m_{01}c^2. \qquad (17)$$

Comparando, (17) e (16), pode-se afirmar que darão valores numéricos aproximados se

$$m_{02}c^2 - m_{01}c^2 = \Delta U_{gás}^{int} \qquad (18)$$

e

$$m_{01} = (M + Nm) \qquad (19)$$

$$m_{02} = (M + Nm + \frac{\Delta U_{gás}^{int}}{c^2}) \qquad (20)$$

em que

$$(M + Nm) \gg \frac{\Delta U_{gás}^{int}}{c^2} \qquad (21)$$

Da expressão (20) conclui-se que, nas condições da 1ª análise, a acção de uma força sobre a massa deformável do gás faz com que a massa do conjunto gás-caixa varie de $\Delta U_{gás}/c^2$. Esta variação resulta da energia interna da massa deformável variar durante o regime transitório, até que a distribuição de massa do gás atinja um novo estado de equilíbrio, imposto pela aceleração no referencial próprio, no referencial do centro de massa. De facto quando a força actua no instante inicial sobre a caixa, não actua imediatamente na totalidade da massa do gás. As forças exercidas pelo gás sobre as duas bases da caixa ainda são iguais porque o gás ainda está distribuido uniformemente. Só quando o gás atinge a distribuição final de equilíbrio é que a força $F$ actua sobre a massa total do gás, "sente" a massa total da caixa-gás. Esta interpretação do conceito de massa é complementar (alternativa) da interpretação electromagnética [14]. Durante esse intervalo de tempo, em que o centro de massa se desloca no interior da caixa, o trabalho realizado pela força $F$ em parte vai para energia cinética da caixa, de translação. Em parte para a energia cinética de translação do gás. E em parte para energia interna do gás. Desta forma, para um corpo extenso, a variação de massa resulta do efeito da aceleração, dado que não existe emissão ou absorção de partículas (como é implicitamente admitido num artigo recentemente publicado - "*the term $c^2$ dm$_0$ keeps track, <u>among other things</u>, of emission and absortion of radiation (or particles)*" - sublinhado nosso [15]). Após o gás atingir a distribuição de equilíbrio o trabalho da força $F$ vai integralmente para energia cinética de translação da caixa e do gás. É como se de um ponto material se tratasse.

## **2. A 2ª Análise**

Consideremos que a caixa está suspensa por um fio num campo gravitacional de aceleração de módulo $g$ (Fig. 3). O valor de $F$, módulo da força exercida pelo fio, é

$$F = (M + Nm)g \qquad (22)$$

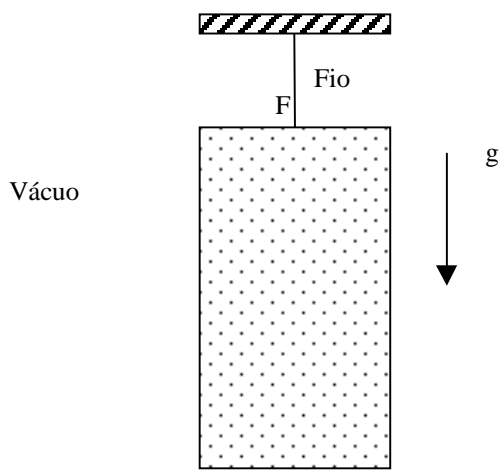

Fig.3 - Uma caixa de massa *M* encontra-se presa por um fio que exerce uma força *(M+Nm)g*. *Nmg* é o peso do gás que se encontra no seu interior.

Neste caso, como é bem conhecido, a força peso de módulo *F*, é equivalente a uma força aplicada no centro de massa. Cortado o fio a aceleração do centro de massa é *g* e o trabalho da força peso é dada por (2). A variação da energia cinética interna do gás é zero. Embora a distribuição da densidade do gás no interior da caixa se altere ficando uniforme, a variação da energia cinética do gás é zero. De facto temos que durante a deformação do gás no interior da caixa as energias cinéticas e potenciais da caixa, $E_{cinCX}$ e $E_{potCX}$ e as energias cinéticas e potenciais do gás, $E_{cing}$ e $E_{potg}$ satisfazem a lei de conservação da energia

$$E_{cinCX} + E_{potCX} + E_{cing} + E_{potg} = const. \qquad (23)$$

ou

$$\Delta E_{cinCX} + \Delta E_{potCX} + \Delta E_{cing} + \Delta E_{potg} = 0. \qquad (24)$$

Como após a deformação as velocidades dos centros de massa da caixa e do gás são iguais à velocidade do centro de massa do conjunto caixa-gás, temos, de (24)

$$\frac{1}{2}Mv_{CM}^2 + \frac{1}{2}Nmv_{CM}^2 + \Delta U_{gás}^{int} = -(\Delta E_{potCX} + \Delta E_{potg}) \qquad (25)$$

De (25), como se demonstra à frente, temos que

$$\Delta U_{gás}^{int} = 0 \qquad (26)$$

(26) resulta da definição de centro de massa e da força *F* estar aplicada no centro de massa.

De facto, o centro de massa define-se por

$$z_{CM} = \frac{Mz_{CMCX} + Nmz_{CMg}}{(M + Nm)} \qquad (27)$$

e, portanto,

$$(M + Nm)g\Delta z_{CM} = Mg\Delta z_{CMCX} + Nmg\Delta z_{CMg} = -\Delta E_{potcx} - \Delta E_{potg} \qquad (28)$$

Como o trabalho da força *F* é dado por

$$(M + Nm)g\Delta z_{CM} = \frac{1}{2}(M + Nm)v_{CM}^2 = -\Delta E_{potcx} - \Delta E_{potg} \qquad (29)$$

dado a força *F* estar aplicada no centro de massa e dado (28). Comparando (29) com (25) temos (26). Concluimos que não há variação de energia interna.

A força peso da caixa actua no centro de massa da caixa. A força peso do gás actua no centro de massa do gás. E a força peso resultante actua no centro de massa do conjunto. Neste caso, embora tenhamos um sistema deformável, o trabalho da resultante das forças exteriores é igual à soma dos trabalhos das forças exteriores que estão aplicadas nos centros de massa dos respectivos sub-sistemas. Contrariamente ao que se passa na 1ª análise, o trabalho da força *F* contribui integralmente para o aumento da energia cinética do conjunto caixa-gás, que, neste sentido, se comporta como se fosse um ponto material.

## Conclusão

O estudo da actuação de forças sobre corpos com estrutura exige duas análises distintas, a que se procedeu no presente artigo, e relacionadas com diferentes condições iniciais, a que se encontram associadas ou não regimes transitórios.

Na primeira análise, Análise I, estudou-se a acção duma força aplicada na fronteira exterior dum corpo. Concluiu-se que quando o corpo se põe em movimento a aceleração duma tal força altera a temperatura no interior do corpo, a sua energia interna e a massa. Isto é consequência de a força no início da sua acção estar simplesmente aplicada na fronteira exterior do corpo, e necessitar de um regime transitório para estender a sua acção ao interior do corpo deformando-o. Quando o corpo pára de se deformar, tudo se passa como se a força actuasse sobre a massa total do corpo, como se fosse uma partícula material.

Porém o campo gravítico exige uma análise diversa, aqui exposta na Análise II. Concluiu-se que neste caso não há variação da temperatura do corpo e da sua energia interna, pois como o corpo sempre esteve mergulhado no campo gravítico, a respectiva força-peso actuou sempre sobre todas e cada uma das suas componentes. Este facto mostra a não necessidade de qualquer regime transitório. Tudo se passa como se de uma partícula material se tratasse.

Estas conclusões contradizem algumas ideias correntes, e.g. as expostas em [1]. Assim o conceito de trabalho mecânico tal como é correntemente leccionado tem mais generalidade do que a sua simples aplicação ao ponto material e ao corpo rígido em translação pura.

Na verdade no caso contemplado na análise II, embora o corpo não seja rígido, o trabalho da resultante está bem definido e é igual à variação da energia cinética do corpo. Este comporta-se como um ponto material. Mas mesmo no caso da Análise I o trabalho da resultante *F* das forças, também está bem definido pois é o produto da força *F* pelo deslocamento do invólucro. E a partir do corpo estar deformado é igual à variação da energia cinética do corpo globalmente considerado, exactamente como num ponto material, ou um corpo rígido em movimento de translação.

Esta abordagem aplica-se ainda a outras situações com as necessárias adaptações, e.g. um corpo a deslizar ao longo dum plano inclinado e no caso dum automóvel.

No caso dum corpo a deslizar num plano inclinado a força de atrito (na qual se deve incluir a resistência do ar) não conduz a um aumento, mas sim a uma diminuição da energia cinética macroscópica do corpo, em relação à que teria se não houvesse atrito, e a um aumento da energia interna do corpo. Trata-se da combinação da Análise I com a Análise II. Quando o corpo inicia o seu movimento a força de atrito é uma força exterior que ao ser transportada para o interior requer um regime transitório. Se o plano for suficientemente extenso e se o regime transitório for suficientemente curto, quando este termina e deixa de haver deformação a energia interna do corpo é constante, passando o corpo a comportar-se como um ponto material. Então o trabalho da força de atrito é em módulo o produto da força de atrito pelo deslocamento do corpo, igual ao deslocamento do centro de massa. Donde se conclui que o deslocamento efectivo só é menor que o deslocamento do centro de massa durante o regime transitório, pela acção da força de atrito, sem necessitar de quaisquer explicações adicionais.

O caso dos automóveis também é interessante. Quando um automóvel está em movimento rectilíneo e uniforme, isso significa que o seu motor produz uma força que contraria a força de resistência do ar, e a de atrito nas rodas. Donde a força global é nula e a variação de energia cinética é igualmente nula. Neste caso a variação da posição do centro de massa do carro, é igual à variação da posição do carro, e tudo se trata como se fosse um ponto material. Não existe deformação. Porém no caso de existir aceleração constante, e após dar-se a deformação, há que ter em conta não só a alteração da energia interna dos pneus, mas também a variação da energia cinética macroscópica de rotação dos pneus. Não se trata de deformação do corpo em regime estacionário [1], mas sim de um problema de massas aceleradas, em rotação com variação de energia interna, problema este a exigir um tratamento próprio.

O presente artigo mostra de forma clara a importância do estudo de regimes transitórios no estudo da mecânica dos corpos extensos. Mostra ainda que este estudo força a inclusão da termodinâmica nos estudos de mecânica tradicional. Esta combinação, que exige uma reflexão profunda por parte dos formadores, não pode ser escamoteada. Embora pareça mais complexa, poderá revelar-se mais interessante aos olhos dos alunos, por evidenciar fenomenologia insuspeita. Este é o caminho formativo a seguir.

## Referências